# Data Hiding Techniques using number decompositions

**Rajdeep Borgohain**
Department of Computer Science and Engineering, Dibrugarh University Institute of Engineering and Technology,
Dibrugarh, India
Email: rajdeepgohain@gmail.com

-------------------------------------------------------------------ABSTRACT-------------------------------------------------------------------
*Data hiding is the art of embedding data into digital media in a way such that the existence of data remains concealed from everyone except the intended recipient. In this paper, we discuss the various Least Significant Bit (LSB) data hiding techniques. We first look at the classical LSB data hiding technique and the method to embed secret data into cover media by bit manipulation. We also take a look at the data hiding technique by bit plane decomposition based on Fibonacci numbers. This method generates more bit planes which allows users to embed more data into the cover image without causing significant distortion. We also discuss the data hiding technique based on bit plane decomposition by prime numbers and natural numbers. These methods are based on mapping the sequence of image bit size to the decomposed bit number to hide the intended information. Finally we present a comparative analysis of these data hiding techniques.*

-----------------------------------------------------------------------------------------------------------------------------------------------

## I. INTRODUCTION

Sharing of digital media is one of the main reasons for the boom of the internet. Users can share between them various information in digital format. But with the increasing threat of security [14] in the cyber world, it has become very important that any important information shared through these digital media be concealed in such a way that only the intended recipient will be able to retrieve the information. In this context data hiding in the form of steganography plays a very important role. The word steganography is derived from Greek which means "Concealed Writing" [15]. The intended data is embedded in an image in such a way that no one but the recipient can extract the hidden message.

For data hiding, the requirements [5] are:
- A medium to hold the hidden information which is known as *cover media.*
- The *secret message* which the sender intends to embed within the cover media.
- A *steganography function* and its inverse.
- An optional key (stego-key) that is used to hide or unhide the data.

For hiding the data, the cover media can be any one of the following digital media [6]:
- Plain Text
- Digital Image
- Audio
- Video
- IP Datagram

The process of hiding data, we first take the cover media which can be of any format as mentioned above. After that we take the secret information which we want to hide in the cover media. The secret message can be in any form, plain text or cipher text or even other image file. The cover media and secret message is then passed through a steganography function and the message is embedded in the cover media. We can also optionally use a key to hide the message and later decode it using the same key [13].

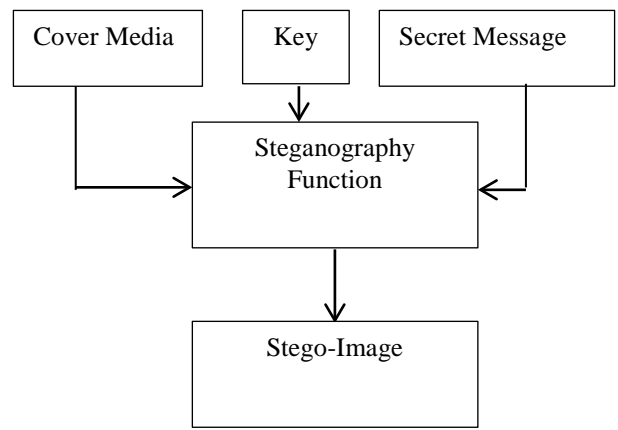

**Figure 1. Steganography Process**

For embedding data in digital media, two domains are generally considered, spatial domain [16, 21] and the transform domain [17, 22]. In the transform domain, the domains considered are Discrete Cosine Transform (DCT) [7], Discrete Fourier Transform (DFT) [23] and Discrete Wavelet Transform (DWT) [8].

Both the spatial and transform embedding scheme has to fulfill three requirements of visibility, robustness and capacity. Visibility is concerned with the fact that human observers should not be able to detect distortions due to the hidden message in the digital media. On the other hand, it is very important that once a secret message is inserted, it becomes impossible to delete or manipulate that message.

This is the requirement for robustness. Moreover, the capacity of the digital media should be kept in consideration while inserting the message. These three factors are dependent on each other and a balance must be maintained between them as increase in one of the factors leads to the decrease in other [9].

Though, there are many data hiding techniques [2, 3, 4], in this paper, we consider the spatial domain of data hiding. We discuss the various data hiding techniques using bit manipulation of the lowest significant bit (LSB). We take a look at how the bit planes can be increased by various number decomposition methods without compromising on the three requirements of visibility, robustness and capacity.

The rest of the paper is organized as follows. In Section 2 we take a look at the classical LSB data hiding technique. Here we discuss how secret text can be embedded in cover image using LSB manipulation. Section 3 of the paper contains the Fibonacci LSB data hiding technique. Section 4 of the paper contains the LSB data hiding technique using decomposition of prime numbers. The approach for decomposition using prime numbers is discussed here. Section 5 contains LSB data hiding by natural numbers. Here the decomposition of bit planes by natural numbers is discussed. In Section 6 we present a comparative analysis between the various LSB data hiding methods. Finally, in Section 7, we present the conclusion for the paper.

## II. CLASSICAL LSB DATA HIDING TECHNIQUE

One of the simplest implementation of data hiding technique is the classical least significant bit data hiding technique [18]. The technique is based on manipulation of the least significant bits of the carrier image to accommodate the hidden message.

The insertion of LSB varies according to the number of bits in an image. For an 8 bit image, the value of $8^{th}$ bit, which is the least significant bit of each pixel, would be modified and the secret message would be embedded [19].

Suppose if we wanted to insert the letter *'A'* into an image. The binary equivalent of *'A'* is 10000001. Now if our sample image of 8 bit has the following pixel values:

10101010 10001010 11111111 010110101 10101010 10101010 10001011 10101111

So, embedding the binary equivalent of *'A'* i.e. 10000001would change the pixel values of our sample image to:

1010101**1** 10001010 1111111**0** 0101101**0** 10101010 10101010 1000101**0** 0101111

In our sample image the least significant bit value of pixel 1, 3, 4 and 7 have been changed.

Similarly, for an image of 24 bit, we need to change the RGB (Red Blue Green) color values. Suppose our sample image has the following pixel values,

[10010101 10101010 11110101]
[10001001 10001110 10000010]
[11110101 01001010 10001011]

So to accommodate our desired character, *'A'*, we have to make the following modifications:

[10010101 10101010 1111010**0**]
[1000100**0** 10001110 10000010]
[1111010**0** 01001010 10001011]

So, basically we have changed the values of 3 bit to accommodate our desired character.

Though, classical data hiding is simple to implement, it is vulnerable to image manipulation [15].

## III. LSB DATA HIDING USING FIBONACCI NUMBERS

Battisti et al. [1] proposed a method of embedding data into digital media by decomposition of Fibonacci number sequence which allowed different bit plane decomposition when compared to the classical LSB scheme.

The Fibonacci sequence, named after Leonardo of Pisa, also known as Fibonacci, is a sequence of numbers in the following integer sequence:

0, 1, 1, 2, 3, 5, 8, 13, 21, 34, 55, 89 …

The Fibonacci numbers are defined by a recurrence relation:

$$F_p(0) = F_p(1) = 1$$

$$F_p(n) = F_p(n-1) + F_p(n-p-1), \quad \forall\ n \geq 2, n \in N$$

The bit planes are now decomposed based on the Fibonacci sequence. The main drawback of this approach is that of redundancy and to counter that and obtain a unique representation, Zeckendorf theorem [20] is used.

To embed the intended message in the cover image, it is decomposed into bit planes by using Fibonacci p-decomposition. The Zeckendorf condition is checked for each bit to be modified. If the condition is fulfilled, the bit is inserted otherwise the bit following it is considered.

## IV. LSB DATA HIDING USING PRIME NUMBERS

LSB data hiding using prime numbers is a data hiding technique proposed by Dey et.al [10] as an improvement over the Fibonacci numbers data hiding technique proposed by Battisti et. al. The main idea of the work was to use the prime number decomposition and generate new set of bit planes and embed information in these newly generated bit planes with minimal distortion.

In this approach, the researchers took an image of *m* bits and increased the number of bit planes to *n,* where the value of *n* was equal or greater than or equal to the number of bit planes of the image. This was achieved by converting the bit planes of the image to another number system using prime numbers as the weighted function. This resulted in the increase of number of bits and consequently it could be used for hiding data in higher bit planes with minimal distortion.

For decomposition, the weight function was defined as:

$$P(0) = 1, P(i) = p_i \forall i \in Z^+, p_i = i^{th} \text{ Prime}$$

In case of any ambiguity, the lexicographically higher number is given preference.

For embedding the data in the image, a number n is chosen in such a way that all possible pixel values in the range of [0, $2^k$ - 1] could be represented using first *n* prime number, so that *n* virtual are achieved after the decomposition.

The value of *n* can be found out by the formula,

$$\sum_{i=0}^{n-1} p_i \geq 2^k - 1$$

After finding the prime numbers, a k-bit to n-bit sequence is mapped, marking all the valid prime number in the system. Now, for each pixel of the image, a virtual bit plane is chosen and secret data is embedded is by modifying the corresponding virtual bit plane by the desired information bit by bit. If the resulting sequence of data after embedding the data matches the k-bit to n-bit mapping, the bits are kept otherwise it is discarded. After insertion of the hidden data, the resultant prime number system is converted back to the original binary system and the stego-image is achieved.

For extraction, each pixel with the hidden information is converted into prime decomposition and from those bit planes, the secret message is extracted. All the bits are combined together to get the final message.

## V. LSB DATA HIDING USING NATURAL NUMBERS

The approach of LSB data hiding by natural numbers was proposed by Dey et.al [11]. In this approach, the researchers proposed data hiding by decomposition of a pixel value in sum of natural numbers. This resulted in generation of more bit planes than the Classical LSB data hiding, Fibonacci LSB data hiding and the Prime number data hiding[10].

For decomposition, the weight function is defined as:

$$W(i) = N(i) = i + 1, \forall i \in Z^+ \cup \{0\}$$

The researchers used the same concept in case of ambiguity which gave higher precedence to lexicographically higher number.

For embedding the data into the k-bit image, a number *n* is chosen in a way such that all pixel values in the range of [0, $2^k$ – 1] could be represented using first n numbers, which resulted in generation of *n* virtual bit planes .

The value of *n* can be found out by the formula,

$$n \geq \frac{-1 \pm \sqrt{2^{k+3} + 9}}{2}$$

After finding the value of n, a k-bit to n-bit map is created and all valid representations in natural numbers system are marked. Now, for each pixel a virtual bit plane is chosen and the secret data is embedded. If the virtual bit plane matches the mapping system, the hidden data is kept otherwise it is discarded. After insertion of the secret message, the natural number system is converted back to its original binary form and the stego-image is achieved.

To extract the message from the stego-image, all the pixels with embedded data bit are converted to the natural number decomposition, and the secret message bits are extracted. Finally, all bits are combined together to get the embedded hidden message.

## VI. COMPARISON OF LSB DATA HIDING TECHNIQUES

In the previous section, we have looked at the various techniques for data hiding using number decomposition. Here, we present a comparative analysis between the various data hiding techniques using number decomposition.

### 6.1 *Technique Used:*

| | |
|---|---|
| Classical LSB | Classical LSB data hiding uses the simplest approach. In classical LSB data hiding, the least significant bit of a pixel is manipulated in order to embed the desired image. |
| Fibonacci Technique | In Fibonacci approach, the bit planes are decomposed so as to generate more bit planes and then the secret message is embedded on following the Zeckendorf theorem. |
| LSB data hiding by prime numbers | In LSB data hiding using prime numbers, the bit planes are decomposed by using sum of prime numbers. After that the secret message is embedded lexicographically. |
| LSB data hiding by natural numbers | In LSB data hiding using natural numbers, the bit planes are decomposed by using sum of natural numbers. Here also the secret message is embedded lexicographically. |

**Table 1. Technique used in LSB data hiding**

## 6.2 *Embedding Technique*

| Classical LSB | Data are inserted in the least significant bit of the cover image. |
|---|---|
| Fibonacci Technique | Technique uses Fibonacci P-sequence for generation of bit planes and data is inserted if it passes the Zeckendorf condition. |
| LSB data hiding by prime numbers | A k-bit to n-bit map is created where the value of n is, $$\sum_{i=0}^{n-1} p_i \geq 2^k - 1$$ Data is inserted bit by bit matching the k-bit to n-bit mapping sequence. |
| LSB data hiding by natural numbers | A k-bit to n-bit map is created where the value of n is, $$n \geq \frac{-1 \pm \sqrt{2^{k+3} + 9}}{2}$$ Data is inserted bit by bit matching the k-bit to n-bit mapping sequence. |

**Table 2. Embedding Techniques used in LSB data hiding**

## 6.3 *Weight Function:*

| Classical LSB | The least significant bit of the image pixel is manipulated. In case of 8 bit image it is the 8[th] bit of each byte. For 24 bit image the RGB color code are changed [12]. |
|---|---|
| Fibonacci Technique | $F_p(0) = F_p(1) = 1$ <br><br> $F_p(n) = F_p(n-1) + F_p(n-p-1),$ <br> $\forall\, n \geq 2, n \in N$ |
| LSB data hiding by prime numbers | $P(0) = 1, P(i) = p_i \,\forall\, i \in Z^+, p_i = i^{th}\, Prime$ |
| LSB data hiding by natural numbers | $W(i) = N(i) = i + 1, \forall\, i \in Z^+ \cup \{0\}$ |

**Table 3. Weight functions**

## 6.4 *Number of bit planes on 8 bit Lena Image:*

| Classical LSB | 8 bit planes using gray scale 8bit Lena image. |
|---|---|
| Fibonacci Technique | 12 bit planes using gray scale 8bit Lena image. |
| LSB data hiding by prime numbers | 15 bit planes using gray scale 8bit Lena image. |
| LSB data hiding by natural numbers | 23 bit planes using gray scale 8bit Lena image. |

**Table 4. Number of bit planes generated.**

## VII. CONCLUSION

In this paper, we looked at the different approaches of the spatial data hiding techniques using Least Significant Bit manipulation. We discussed about the classical LSB data hiding technique which is based on manipulation of the least significant bit. We took a further look at the different number decomposition techniques by which more bit planes could be generated to conceal more data without causing significant distortion. By comparative analysis, we could see that among the data hiding techniques, decomposition using natural numbers yielded the highest number of bit planes which was 23. This was followed by 15, 12 and 8 bit of Prime number technique [11], Fibonacci Technique and Classical LSB technique respectively. Finally we can say that in this era of internet where there is increasing use of digital format to send valuable information and data, steganography will play a very important role in the days to come.